\documentclass[11pt]{article}

\usepackage{fullpage, cite, dsfont, graphicx, amsmath, authblk}

\input{Qcircuit}

\title{Fast quantum computation at arbitrarily low energy}

\author[1,2]{Stephen P. Jordan\footnote{\texttt{stephen.jordan@nist.gov}}}

\affil[1]{\small{National Institute of Standards and Technology, Gaithersburg, MD}}
\affil[2]{\small{Joint Center for Quantum Information and Computer Science (QuICS), University~of~Maryland,~College~Park,~MD}}

\date{}

\begin{document}

\maketitle

\begin{abstract}
One version of the energy-time uncertainty principle states that the minimum time $T_{\perp}$ for a quantum system to evolve from a given state to any orthogonal state is $h/(4 \Delta E)$ where $\Delta E$ is the energy uncertainty. A related bound called the Margolus-Levitin theorem states that $T_{\perp} \geq h/(2 \langle E \rangle)$ where $\langle E \rangle$ is the expectation value of energy and the ground energy is taken to be zero. Many subsequent works have interpreted $T_{\perp}$ as defining a minimal time for an elementary computational operation and correspondingly a fundamental limit on clock speed determined by a system's energy. Here we present local time-independent Hamiltonians in which computational clock speed becomes arbitrarily large relative to $\langle E \rangle$ and $\Delta E$ as the number of computational steps goes to infinity. We argue that energy considerations alone are not sufficient to obtain an upper bound on computational speed, and that additional physical assumptions such as limits to information density and information transmission speed are necessary to obtain such a bound.
\end{abstract}

\newcommand{\id}{\mathds{1}}
\newcommand{\eq}[1]{(\ref{#1})}
\renewcommand{\th}{^\mathrm{th}}
\newcommand{\sect}[1]{\S \ref{#1}}

\section{Introduction}
\label{sec:intro}

It is frequently argued that energy places a fundamental limit on the speed of computation, even if the computation is performed reversibly. Planck's constant provides a conversion factor between energy and frequency, and on dimensional grounds one might expect that a quantum system at energy scale $E$ has maximum computational clock speed of order $E/h$. More quantitatively, for a quantum system with energy uncertainty $\Delta E$, the time $T_{\perp}$ to evolve to an orthogonal quantum state obeys the bound \cite{MT45, F73, AA90, V92, U93}
\begin{equation}
T_{\perp} \geq \frac{h}{4 \Delta E}.
\end{equation}
Several related energetic lower bounds on $T_{\perp}$ have been proven \cite{P95, ML98, P93, BCM96, GLM03, LT09}. In particular, the Margolus-Levitin theorem \cite{ML98} shows that for a state with energy expectation $\langle E \rangle$, evolving according to a time-independent Hamiltonian with zero ground energy,
\begin{equation}
T_{\perp} \geq \frac{h}{2 \langle E \rangle}.
\end{equation}
See \cite{Frey16} for a recent review of such quantum speed limits and their applications. Many estimates of computational capacity of physical systems have their starting point in the assumption that $T_{\perp}^{-1}$ can be interpreted as a maximum computational clock speed \cite{Lloyd00, M14, LN07, L02, F02, LT09, M03, BRSSZ16a, BRSSZ16b, Hsu06, Hsu07, GPP06, Frank05, Ng08}. Related, but distinct, arguments for limits on computational speed related to energy are given in \cite{B62, B82, B81, L82}. Another relationship between the energy-time uncertainty principle and computational complexity, which is not based on associating $T_{\perp}$ with the time necessary for a logical operation, was recently given in \cite{AA16}.

For several reasons, equating $T_{\perp}^{-1}$ with a maximum computational clock speed seems quite plausible. In classical computers, each logic gate brings the system to an orthogonal state, and in quantum circuits, each logic gate typically\footnote{As a rough argument, one could model a typical state at an intermediate step of a quantum computation by a Haar random state on $n$ qubits. In this case, the root-mean-square inner product between the states before and after a one-qubit gate has been performed is $1/\sqrt{2^n}$.} brings the system to a near-orthogonal state. $T_{\perp}$ is the minimum time to flip a qubit from $\ket{0}$ to $\ket{1}$ and it would seem surprising to achieve $G$ logical gates in time less than $G T_{\perp}$. Furthermore, energy-time uncertainty principles suggest that the uncertainty of the timing of a logic operation scales as $1/\Delta E$, and if the time between elementary logic operations is shorter than this then one would expect their ordering to be uncertain, thereby ruining the computation. (For a contrary conjecture, perhaps presaging the present work, see \cite{BL85}.) 

However, we here show that these obstacles can be evaded. We give explicit constructions of quantum time-evolutions using time-independent Hamiltonians, which simulate the operation of a quantum circuit of $G$ elementary gates, while traversing only a constant number of orthogonal states, independent of $G$. The Hamiltonians achieving this involve only 4-qubit interactions and can be made spatially local in two dimensions. The ratio of the computational clock speed to $\Delta E$ is unbounded, specifically growing linearly with $G$. The same holds for $\langle E \rangle$. In other words, rather counterintuitively, the total time needed to simulate $G$ gates is on the same order as the minimum required time to flip a single bit. We conclude that energy alone does not present a fundamental limit to computational speed; to obtain such a limit one must invoke additional assumptions such as a limit to the spatial density at which qubits can be packed. The argument based on energy-time uncertainty principles is evaded because, although the uncertainty of the timing of each individual logic operation is very large in our construction, these timings are all correlated, leaving no ambiguity as to the ordering of the operations.

\section{Basic Construction}
\label{sec:basic}

Here we show how to simulate an arbitrary quantum circuit of $G$ gates by evolving for some time $T$ according to a time-independent 4-local Hamiltonian. The computational clock speed achieved $f_{\mathrm{clock}}= G/T$ is such that the ratios  $f_{\mathrm{clock}}/\langle E \rangle$ and $f_{\mathrm{clock}}/\Delta E$ both diverge linearly with $G$. This construction thus suffices to serve as a counterexample for the conjectures that computational clock speed is limited to some maximum rate proportional to either the energy uncertainty or the energy above the ground state. To those familiar with Feynman-Kitaev clock Hamiltonians, the essential idea of the construction can be concisely expressed: it is to initialize a Feynman-Kitaev-type Hamiltonian with a wavepacket whose breadth is comparable to the number of gates $G$. Such broad wavepackets have small expected energy and low energy uncertainty. Furthermore, they maintain large overlap with previous states as they propagate, and thus the number of orthogonal states traversed during the time evolution is $O(1)$ even as the number of computational steps $G$ is increased. In the remainder of this section, the construction is given in detail without assuming prior familiarity with the Feynman-Kitaev construction. Note that here we are using the Feynman-Kitaev Hamiltonian to execute quantum computation ``ballistically'' as in Feynman's original construction \cite{F85} not adiabatically as in \cite{adiabatic}.

Consider a quantum circuit $U$ consisting of a sequence of 2-qubit gates $U = U_G U_{G-1} \ldots U_1$ acting on $n$ qubits. For a given initial state $\ket{h_0}$ the circuit proceeds through the states $\ket{h_0} \to \ket{h_1} \to \ldots \to \ket{h_G}$ where
\begin{equation}
\begin{array}{rcl}
\ket{h_1} & = & U_1 \ket{h_0} \\
\ket{h_2} & = & U_2 U_1 \ket{h_0} \\
             & \vdots & \\
\ket{h_G} & = & U_G \ldots U_1 \ket{h_0}.
\end{array}
\label{history}
\end{equation}
The standard Feynman-Kitaev clock Hamiltonian \cite{F85, KSV02} acts on a register of $n$ computational qubits, together with a clock register, as
\begin{equation}
\label{hfk}
H^{\mathrm{FK}} = \sum_{x = 1}^G \left(-U_x \otimes \ket{x}\bra{x-1} - U_x^\dag \otimes \ket{x-1}\bra{x} + \id \otimes \ket{x}\bra{x} + \id \otimes \ket{x-1} \bra{x-1} \right),
\end{equation}
where $\id$ denotes the identity matrix, the first tensor factor represents the computational qubits, and the second tensor factor represents the clock register. The subspace
\begin{equation}
\mathcal{C} = \mathrm{span} \{ \ket{h_0} \ket{0}, \ket{h_1} \ket{1}, \ldots, \ket{h_G} \ket{G} \}
\end{equation}
is preserved by $H^{\mathrm{FK}}$ and the block of $H^{\mathrm{FK}}$ acting on this subspace looks like
\begin{equation}
\left. H^{\mathrm{FK}} \right|_\mathcal{C} = \left[ \begin{array}{rrrrr}
1 & -1 &     &        &   \\
-1& 2  & -1  &        &   \\
  & -1 &  2  & -1     &   \\
  &    &     & \ddots &   \\
  &    &     &    -1  & 1 \\
\end{array} \right]
\end{equation}
with all other entries zero. In other words, $\left. H^{\mathrm{FK}} \right|_{\mathcal{C}}$ is the discretized second derivative on a one-dimensional lattice. Formally, if we think of $\left. H^{\mathrm{FK}} \right|_\mathcal{C}$ as acting on a discretization of the unit interval $[0,1]$ one has
\begin{equation}
\lim_{G \to \infty} G^2 \left. H^{\mathrm{FK}} \right|_{\mathcal{C}}  = -\frac{\partial^2}{\partial x^2}.
\end{equation}
Consequently, the dynamics induced by $H^{\mathrm{FK}}$ on the computational subspace $\mathcal{C}$ can be intuitively thought of as the dynamics of a free particle on a line. A state $\sum_x \psi(x) \ket{h_x} \ket{x}$ will evolve according to\footnote{For notational simplicity, we use units where $\hbar = 1$ here and throughout the remainder of this paper.}
\begin{equation}
\label{line}
i \frac{\partial}{\partial t} \psi(x,t) = H^{\mathrm{FK}} \psi(x,t) \simeq -\frac{1}{G^2} \frac{\partial^2}{\partial x^2} \psi(x,t).
\end{equation}
To perform computation, one can prepare an initial wavepacket at small $x$ with momentum in the direction of increasing $x$. Under Schr{\"o}dinger time evolution according to \eq{line} this wavepacket will propagate to larger $x$ and broaden. Eventually, the wavepacket will reach the end of the line, and consist of a superposition with nonzero amplitude on $\ket{\psi_G}$, which is the output of the original quantum circuit.

One might be concerned about two apparent problems with the Feynman-Kitaev construction. Firstly, the initial wavepacket must have some width large compared to the lattice spacing in order for the continuum approximation to be valid. Thus it has support not only on the initial state $\ket{h_0} \otimes \ket{0}$ but also on states $\ket{h_1} \otimes \ket{1}, \ket{h_2} \otimes \ket{2},\ldots$ in which some gates have already been applied. This seems like cheating; we are seeking to simulate the computation performed by the quantum circuit $U$, but we start with an initial state in which some of this computation has already been done. This problem is easily solved by padding the circuit $U$ with sufficiently many initial identity gates so that the initial wavepacket only has support on states in which no nontrivial gates have been applied. Secondly, upon measuring the final state, the probability of obtaining the desired outcome $\ket{h_G} \otimes \ket{G}$ will be much smaller than one because the final wavepacket has nonzero amplitude on states $\ket{h_{G-1}} \otimes \ket{G-1}, \ket{h_{G-2}} \otimes \ket{G-2},\ldots$ in which not all of the gates have yet been applied. This second problem can be similarly solved by padding the circuit with sufficiently many identity gates at the end.

So far, we have not specified the physical implementation of the clock register. We have simply labeled an orthonormal basis $\ket{0}, \ket{1}, \ldots, \ket{G}$ for its Hilbert space. We can implement this using $G+1$ qubits in the following encoding.
\begin{equation}
\begin{array}{rcl}
\ket{0} & \mapsto & \ket{100\ldots0} \\
\ket{1} & \mapsto & \ket{010\ldots0} \\
        & \vdots  & \\
\ket{G} & \mapsto & \ket{000\ldots1} 
\end{array}
\label{pulseclock}
\end{equation}
This encoding is inefficient in the sense that only $\log_2 (G+1)$ qubits are actually needed to store numbers in the range $\{0,2,\ldots,G\}$. However, the advantage of this encoding is that the operators such as $\ket{x-1}\bra{x}$ appearing $H^{\mathrm{FK}}$ only act on two qubits, namely qubits $x$ and $x-1$. Thus, if the original circuit $U$ is constructed from a universal gate set of one-qubit and two-qubit quantum gates, then $H^{\mathrm{FK}}$ is a 4-local Hamiltonian.

\begin{figure}[htb]
\begin{center}
\includegraphics[width=0.7\textwidth]{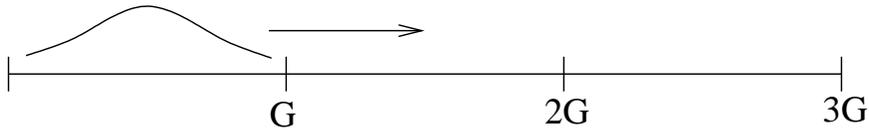}
\caption{The circuit $U = U_G \ldots U_1$ is padded by $G$ identity gates at the beginning and end. The initial wavepacket has width of order $G$ and support on the initial identity gates. The wavepacket then propagates rightward, ending with its support on the final segment of $G$ identity gates.}
\label{fig:broad}
\end{center}
\end{figure}

Let's first treat broadening of the wavepacket as negligible and take the width of the wavepacket to be $G$, as illustrated in figure \ref{fig:broad}. (We defer analysis of broadening to \sect{sec:dispersion}.) The original circuit $U=U_G \ldots U_1$ should be padded with $G$ identity gates at the beginning and $G$ identity gates at the end, yielding a circuit $U' = U_{3G}' U_{3G-1}' \ldots U_1'$ with a total of $3G$ gates and
\begin{equation}
U_x' = \left\{ \begin{array}{rl}
\id & 1 \leq x \leq G \\
U_{x-G} & G+1 \leq x \leq 2G \\
\id & 2G+1 \leq x \leq 3G.
\end{array} \right.
\end{equation}
The initial state can be chosen as a wavepacket with width on the order of $G$ and support only on the initial $G$ identity gates. Let $\ket{h_x'} = U_x' U_{x-1}' \ldots U_1' \ket{h_0}$ while retaining the notation $\ket{h_x}$ defined in \eq{history} for the $x\th$ state obtained by the original unpadded circuit. Thus the initial wavepacket is:
\begin{equation}
\sum_{x=0}^G \psi(x) \ket{h_x'} \ket{x} = \ket{h_0} \left( \sum_{x=0}^G \psi(x) \ket{x} \right).
\end{equation}
That is, the computational register is initialized to the same state that the original circuit starts with, which could be taken as $\ket{0\ldots0}$ without loss of generality\footnote{Quantum circuits acting on the all zeros string are computationally universal because any classical input string can be hard-coded via initial NOT gates.}. The wavepacket $\psi(x)$ should be chosen with rightward momentum. Then it will propagate down the line, yielding a final state of
\begin{equation}
\sum_{x=0}^G \psi(x) \ket{h_{x+2G}'} \ket{x+2G} = \ket{h_G} \left( \sum_{x=0}^G \psi(x) \ket{x+2G} \right).
\end{equation}
Thus, in the final state, the computational register contains the output of the original circuit. 

From figure \ref{fig:broad} one sees that, as the wavepacket propagates, it passes through essentially three orthogonal states: the initial state supported on $0$ through $G$, the state supported on $G$ through $2G$, and the final state supported on $2G$ through $3G$. The energy uncertainty in a state of spatial width $G$ for the Hamiltonian $H^{\mathrm{FK}}$ scales as $\Delta E \sim 1/G^2$. The ground energy of the Feynman-Kitaev Hamiltonian is zero. One can verify this by noting that the state 
\begin{equation}
\frac{1}{\sqrt{G+1}} \sum_{x=0}^G \ket{h_x} \ket{x}
\end{equation}
is an energy-zero eigenvector of $H^{\mathrm{FK}}$ and that, by \eq{hfk} and \eq{pulseclock}, $H^{\mathrm{FK}}$ is positive semidefinite since it is a sum of $G$ positive-semidefinite terms. 

Let $\mathcal{G}$ denote the number of gates in the padded quantum circuit. (If the wavepacket did not spread then we could take $\mathcal{G} = 3G$ while still ensuring that the final superposition is only over states in which the computation is finished. In the present construction we will take $\mathcal{G}$ slightly larger than $3G$, but still only linear in $G$, as discussed in \sect{sec:dispersion}.) Examining \eq{line} with $G \mapsto \mathcal{G}$, one sees that, in the limit of large $\mathcal{G}$, the wavepacket propagates like a nonrelativistic particle of mass $m = \mathcal{G}^2/2$ on the unit interval. The velocity of such a particle (\emph{i.e.} the group velocity of the wavepacket) is $\sim p/m \sim p/\mathcal{G}^2$, where $p$ is the momentum of the wavepacket. The time needed to propagate to the end of the line is therefore $T \sim \mathcal{G}^2/p$. During this time, $G$ gates are simulated, so the clock speed is $f_{\mathrm{clock}} \sim G/T \sim Gp/\mathcal{G}^2$. As we show in \sect{sec:dispersion}, it suffices to choose $\mathcal{G}$ to be a constant multiple of $G$. Thus, $f_{\mathrm{clock}} \sim p/G$. The expectation value of the energy in the wavepacket state is $\langle E \rangle = p^2/2m \sim p^2/G^2$. Hence, by choosing $p = O(1)$ we can achieve a ratio of clock speed to energy of $f_{\mathrm{clock}}/\langle E \rangle \sim G$. As mentioned above, with wavepacket of width of order $G$, the energy uncertainty is of order $1/G^2$. Hence, the ratio of clock speed to energy uncertainty is also of order $G$. 

\section{Dispersion}
\label{sec:dispersion}

Suppose the initial state is a Gaussian wavepacket
\begin{equation}
\psi(x,0) = \eta \exp \left[ - \frac{x^2}{2 \sigma_x^2} + i p_0 x \right]
\end{equation}
where $\eta$ is a normalization constant. Under Schr{\"o}dinger's equation, $i \frac{\partial \psi}{\partial t} = -\frac{1}{2m} \frac{\partial^2}{\partial x^2}$, this evolves to\footnote{An easy way to obtain this is by going to Fourier space, where the Hamiltonian is diagonal.}
\begin{equation}
\psi(x,t) = \eta(t) \exp \left[ - \frac{(x - x_0(t))^2}{2 \sigma_x(t)^2} + i \phi(x,t) \right]
\end{equation}
where $\phi(x,t)$ is a (real) phase, $\eta(t)$ is a normalization factor, and
\begin{eqnarray}
x_0(t) & = & \frac{p_0 t}{m} \\
\sigma_x(t) & = & \sqrt{ \sigma_x^2 + \left( \frac{t}{m \sigma_x} \right)^2}.
\label{breadth}
\end{eqnarray}

In the units being used, where the length of the line is 1 and the lattice spacing is $1/\mathcal{G}$, we have, in the continuum limit (\emph{i.e.} $\mathcal{G} \to \infty$), an effective particle mass of
\begin{equation}
m = \frac{\mathcal{G}^2}{2}
\end{equation}
and a propagation velocity
\begin{equation}
\label{velocity}
v = \frac{p_0}{m} = \frac{2 p_0}{\mathcal{G}^2}
\end{equation}
Thus, the time to propagate down the line is $T \sim 1/v$ and $\sigma_x(T)$ is O(1). In other words, the final superposition has width only a constant factor larger than the initial superposition. To obtain a wavepacket with finite support, one can truncate the Gaussian wavepacket at some multiple of $\sigma_x$ away from the mean. That is,
\begin{equation}
\psi_{\mathrm{trunc}}(x,0) = \left\{ \begin{array}{cl} 
\eta \exp \left[ - \frac{x^2}{2 \sigma_x^2} + i p_0 x \right] & |x - x_0| \leq c \sigma_x \\
0 & \textrm{otherwise.} \end{array} \right.
\end{equation}
The inner product between this wavepacket and the untruncated Gaussian becomes exponentially close to one as the multiple $c$ is increased. By unitarity, the inner product between the ideal and truncated final states will be equal to the inner product between the ideal and truncated initial states. Because initial and final $\sigma_x$ are $O(1)$, the initial truncated wavefunction has support only on $O(G)$ clock values and can be completely accommodated by padding the circuit with $O(G)$ initial identity gates. The final state will, to an exponentially good approximation for large $c$, have all its amplitude within a range of $O(G)$ clock values and can also be accommodated by padding with $O(G)$ final identity gates.

As a concrete example, suppose we start with a circuit of $G$ gates, and we pad it with $2G$ initial identity gates and $2G$ final identity gates. Thus, $\mathcal{G} = 5G$. We set the initial state to be a Gaussian superposition of width $\frac{G}{4}$ and mean $G$. Thus, we can take $c = 4$ and have zero amplitude outside the initial pad, whose width is $2G$. This ensures that the initial state has zero amplitude for any of the computational gates to be already computed. An initial superposition of width $\frac{G}{4}$ means, in units where the length of the line is one, $\sigma_x = \frac{1}{20}$. We want the wavepacket to propagate from the middle of the initial pad to the middle of the final pad. Thus the total distance to propagate is $\frac{3}{5}$. By \eq{velocity} this will take time $T = \frac{3 \mathcal{G}^2}{10 p_0}$. Choosing $p_0 = 240$ (with corresponding energy $\frac{p_0^2}{\mathcal{G}^2}$) yields, by \eq{breadth}, a final wavepacket width of $\sigma_x(T) = \frac{\sqrt{2}}{20}$. Measurement of the clock register in the final wavepacket will yield a Gaussian probability distribution with mean $4G$ and standard deviation\footnote{The 2 comes from squaring the amplitudes. The $5$ comes from $\mathcal{G} = 5G$.} $2 \times 5G \sigma_x(T) = \frac{1}{\sqrt{2}} G$. Almost all the probability thus lies in the final pad of identity gates covering clock values $3G$ to $5G$.  Outcomes in which the clock register has value less than $3G$ correspond to events at least $\sqrt{2}$ standard deviations below the mean, which have total probability about $8\%$. Hence, including the errors from the truncation of the initial wavepacket, one finds that, upon measuring the final state, the computational register will with probability at least $88\%$ be in a state where all $G$ gates have been computed.

\section{Distance Traversed Through Hilbert Space}

The question whether the number of orthogonal states traversed in a computation is a resource bearing on computational capacity is perhaps interesting independent of connection to energy. It could be asked in the context of computational models where energy may not play a manifest role, such as quantum cellular automata, quantum Turing machines, and the quantum circuit model. To ask this question in a precise way, we first note that the number of orthogonal states traversed during a computation is not a well-behaved metric. Instead, for a discrete-time quantum computation, such as the quantum circuit model, we can formalize the intuitive notion of distance traversed through Hilbert space as
\begin{equation}
\mathcal{L} = \sum_{x = 1}^G \left\| \ket{h_x} - \ket{h_{x-1}} \right\|
\end{equation}
where $\ket{h_x}$ is the quantum state obtained after the first $x$ gates have been applied, as in \eq{history}. The continuum analogue of this distance is
\begin{equation}
\mathcal{L}(\ket{\psi(t)};t_2,t_1) = \int_{t_1}^{t_2} dt \left\| \frac{d \ket{\psi(t)}}{dt} \right\|,
\end{equation}
which can be applied to Hamiltonian-based models of quantum computation. Note that the continuum version of $\mathcal{L}$ remains invariant if we multiply the speed at which we traverse the path through Hilbert space by some factor, as is fitting for a metric of distance.

The construction of \sect{sec:basic} demonstrates that it is possible to simulate $G$ gates while keeping $\mathcal{L}=O(1)$. One way to see this is by the following calculation.
\begin{eqnarray}
\mathcal{L} & = & \int_0^T dt \left\| \frac{d}{dt} \ket{\psi} \right\| \\
& = & T \left\| H \ket{\psi} \right\| \\
& = & T \sqrt{\bra{\psi} H^2 \ket{\psi}} \\
& = & T \sqrt{(\Delta E)^2 + \langle E \rangle^2} \\
& = & O(1)
\end{eqnarray}
The last line follows from the results of \sect{sec:basic} stating that $T = O(G^2)$, $\Delta E = O(1/G^2)$, and $\langle E \rangle = O(1/G^2)$. 

\section{Dispersionless Discretization}
\label{sec:dispersionless}

In the construction of \sect{sec:basic}, the dispersion relation is quadratic and therefore the wavepacket spreads as it propagates. Here we give an alternative construction which avoids this complication by modifying the Feynman-Kitaev Hamiltonian to yield a linear dispersion. This modified construction achieves constant clock speed with energy uncertainty scaling as $\Delta E \sim 1/G$. This is thus a faster method of computation than the basic construction, which achieves $O(1/G)$ clock speed at $O(1/G^2)$ energy uncertainty\footnote{These speeds and energy scales are determined by our choice of normalization of the Hamiltonian, but this choice is not arbitrary. Physically, one expects the individual 4-local terms to have $O(1)$ norm.}. Furthermore, in the linear-dispersion construction $\lim_{G \to \infty} \mathcal{L}$ is easy to calculate in complete quantitative detail. However, it uses wavepackets whose energy is not close to the ground energy. Depending on context this may or may not be relevant, which is why both constructions are presented in this manuscript. Roughly speaking, linear dispersion is achieved by discretizing a one-dimensional analogue of Dirac's equation rather than discretizing the one-dimensional Schr{\"o}dinger equation. Similar ideas have been used previously in \cite{B82, SB13}. 

Consider the Hamiltonian
\begin{equation}
H = v \left[ \begin{array}{cc}
0 & - \frac{\partial}{\partial x} \\
\frac{\partial}{\partial x} & 0 \end{array} \right]
\end{equation}
where $v$ is a ``velocity.'' This is Hermitian because $\frac{\partial}{\partial x}$ is antihermitian. Schr{\"o}dinger's equation then reads
\begin{equation}
\frac{\partial}{\partial t} \left[ \begin{array}{c} \Phi \\ \Psi \end{array} \right] = -i v \left[ \begin{array}{cc} 0 & - \frac{\partial}{\partial x} \\ \frac{\partial}{\partial x} & 0 \end{array} \right] \left[ \begin{array}{c} \Phi \\ \Psi \end{array} \right].
\end{equation}
Consequently,
\begin{equation}
\frac{\partial^2}{\partial t^2} \left[ \begin{array}{c} \Phi \\ \Psi \end{array} \right] = v^2 \left[ \begin{array}{cc} \frac{\partial^2}{\partial x^2} & 0 \\ 0 & \frac{\partial^2}{\partial x^2} \end{array} \right] \left[ \begin{array}{c} \Phi \\ \Psi \end{array} \right].
\end{equation}
That is, $\Phi$ and $\Psi$ each obey the one-dimensional wave equation. A solution to the Schr{\"o}dinger equation is therefore
\begin{equation}
\left[ \begin{array}{c} \Phi \\ \Psi \end{array} \right] = \left[ \begin{array}{cc} w(x-vt) \\ iw(x-vt) \end{array} \right]
\end{equation}
for any function $w$, as can easily be verified. $w$ describes the shape of a wavepacket that rigidly propagates in the positive-$x$ direction without distortion.

We can now discretize $\frac{\partial}{\partial x}$ and $-\frac{\partial}{\partial x}$ using finite differences while maintaining Hermiticity. To achieve this, we use a forward difference to discretize $\frac{\partial}{\partial x}$ and a backward difference to discretize $-\frac{\partial}{\partial x}$ as illustrated by the following example on a lattice of four sites.
\begin{equation}
\label{symplect}
H = \left[ \begin{array}{rrrrrrrr}
0 & 0 & 0 & 0 & 1 & -1 & 0 & 0 \\
0 & 0 & 0 & 0 & 0 & 1 & -1 & 0 \\
0 & 0 & 0 & 0 & 0 & 0 & 1 & -1 \\
0 & 0 & 0 & 0 & 0 & 0 & 0 & 1 \\
1 & 0 & 0 & 0 & 0 & 0 & 0 & 0 \\
-1 & 1 & 0 & 0 & 0 & 0 & 0 & 0 \\
0 & -1 & 1 & 0 & 0 & 0 & 0 & 0 \\
0 & 0 & -1 & 1 & 0 & 0 & 0 & 0
\end{array} \right]
\end{equation}
Using $\ket{\Phi_0}, \ldots, \ket{\Phi_{\mathcal{G}/2}}$ as a basis for the $\Phi$ subspace and $\ket{\Psi_0}, \ldots, \ket{\Psi_{\mathcal{G}/2}}$ for the $\Psi$ subspace, one has
\begin{equation}
\begin{array}{lcr}
\bra{\Phi_x} H \ket{\Psi_x} & = & 1 \\
\bra{\Phi_x} H \ket{\Psi_{x-1}} & = & -1
\end{array}
\end{equation}
with all other matrix elements zero other than the Hermitian conjugates of the above.

The linear-dispersion Feynman-Kitaev computational Hamiltonian corresponding to this discretization is
\begin{equation}
\begin{array}{rl} H_{\mathrm{lin}} = \displaystyle{\sum_{x = 1}^{\mathcal{G}/2}} \Big[ & U_{2x}^\dag \otimes \ket{\Phi_x} \bra{\Psi_x} + U_{2x} \otimes \ket{\Psi_x} \bra{\Phi_x} \\
 & -  U_{2x-1} \otimes \ket{\Phi_x} \bra{\Psi_{x-1}} - U^\dag_{2x-1} \otimes \ket{\Psi_{x-1}} \bra{\Phi_x} \Big].
\end{array}
\end{equation}
We can use $\mathcal{G}+1$ qubits to encode the clock state analogously to \eq{pulseclock}. That is,
\begin{eqnarray*}
\ket{\Phi_0} & = & \ket{10000\ldots0} \\
\ket{\Psi_0} & = & \ket{01000\ldots0} \\
\ket{\Phi_1} & = & \ket{00100\ldots0} \\
\ket{\Psi_1} & = & \ket{00010\ldots0} \\
& \vdots & 
\end{eqnarray*}
Then, the clock transitions are again 2-local operators. Thus, if the original circuit is built from gates that each act on at most two qubits, $H_{\mathrm{lin}}$ is a 4-local Hamiltonian.

Using this encoding, $H_{\mathrm{lin}}$ is an operator acting on a $2^{n+\mathcal{G}}$-dimensional Hilbert space. However, as with the Feynman-Kitaev Hamiltonian, the subspace $\mathcal{C}$ is preserved by this Hamiltonian, and within $\mathcal{C}$, $H_{\mathrm{lin}}$ acts as illustrated in \eq{symplect}. Discretizing $\frac{\partial}{\partial x}$ by a forward difference on a lattice of spacing $a$ corresponds to
\[
\frac{1}{a} \left[ \begin{array}{rrrr}
-1&  1 &    & \\
  & -1 &  1 & \\
  &    & -1 & 1 \\
  &    &    & \ddots \\ 
\end{array} \right].
\]
So, if we think of the clock register as discretizing the unit interval, the corresponding wave propagation speed $v$ is $1/\mathcal{G}$, up to higher order corrections in $1/\mathcal{G}$.

With $H_{\mathrm{lin}}$ we can achieve arbitrary length computations with constant-length paths through Hilbert space just as in the Feynman-Kitaev example, but now the evolution of the wavepacket is simpler and cleaner to analyze. We pad the circuit with $G$ initial identity gates and $G$ final identity gates. Then, we prepare the initial wavepacket state
\begin{equation}
\ket{0}^{\otimes n} \otimes \frac{1}{\sqrt{2}} \sum_{x=0}^{G/2} \left[  w\left( \frac{x}{G} \right) \ket{\Phi_x} + i w \left( \frac{x}{G} \right) \ket{\Psi_x} \right]
\end{equation}
where $w$ is a normalized wavepacket and, without loss of generality, we take $\ket{0}^{\otimes n}$ to be the initial state of the computational register. 

We can now compute $\mathcal{L}$ for this construction in the limit $G \to \infty$. In this continuum limit we have
\begin{equation}
\label{contin}
\ket{\psi(t)} = \frac{1}{\sqrt{2}} \int_0^1 dx \ w(x-vt) \ket{x_\Phi} + \frac{i}{\sqrt{2}} \int_0^1 dx \ w(x-vt) \ket{x_\Psi}.
\end{equation}
where $\ket{x_\Phi}$ and $\ket{x_\Psi}$ are the continuum analogues of $\ket{h'_{2x}}\ket{\Phi_x}$ and $\ket{h'_{2x+1}}\ket{\Psi_x}$, respectively. Here we have chosen the normalization so that
\begin{equation}
\int_0^1 w(x)^2 = 1.
\end{equation}
Thus, one finds
\begin{equation}
\left\| \frac{d}{dt} \ket{\psi} \right\| = \sqrt{2} v \int_0^1 dx |w'(x)|^2,
\end{equation}
where $w'$ denotes the derivative of $w$. In our construction padded with identity gates the propagation velocity is $v \simeq \frac{1}{3G}$ and the total duration is $T \simeq 3G$. So
\begin{equation}
\mathcal{L} = \int_0^{3G} dt \frac{\sqrt{2}}{3G} \int_0^1 dx |w'(x)|^2 = \sqrt{2} \int_0^1 dx |w'(x)|^2.
\end{equation}
As a concrete example of a smooth normalized wavepacket $w(x)$ with support only on $0 \leq x \leq \frac{1}{3}$, one could choose
\begin{equation}
w(x) = \left\{ \begin{array}{cl} \sqrt{2} \left( 1 - \cos(6 \pi x) \right) & 0 \leq x \leq \frac{1}{3} \\
0 & \textrm{otherwise}. \end{array} \right.
\end{equation}
In this case, one finds by straightforward calculation that
\begin{equation}
\label{lval}
\mathcal{L} = 12 \sqrt{2} \pi^2.
\end{equation}

Next we can compute $\Delta E$. By \eq{contin}
\begin{eqnarray}
\bra{\psi} H \ket{\psi} & = & i \bra{\psi} \frac{d}{dt} \ket{\psi} \\
 & = & i v \int_0^1 dx w(x-vt) w'(x-vt)\\
 & = & 0.
\end{eqnarray}
Thus,
\begin{eqnarray}
\Delta E & = & \sqrt{\bra{\psi} H^2 \ket{\psi}} \\
 & = & \| H \ket{\psi} \| \\
 & = & \left\| \frac{d}{dt} \ket{\psi} \right\| \\
 & = & \frac{\mathcal{L}}{3G} \\
 & = & \frac{4 \sqrt{2} \pi^2}{G}
\end{eqnarray}
by \eq{lval}.

\section{Spatially Local Construction}
\label{sec:local}

The constructions of \sect{sec:basic} and \sect{sec:dispersionless} are perhaps slightly unphysical in that, although the Hamiltonians are local in the sense of involving only 4-qubit interactions, they are not spatially local. In this section we describe how to use an idea from \cite{Nagaj12} to modify the constructions so that they become spatially local in two dimensions. A detailed illustrative example is given in appendix \ref{layers}.

We arrange the gates of the original circuit into ``layers'' such that the gates within each layer act on distinct subsets of the qubits. The number of layers $D$ is called the circuit depth. Correspondingly, we have a sequence $\ket{l_0},\ket{l_1}, \ldots, \ket{l_D}$ where $\ket{l_j}$ is the state of the qubits after $j$ of the layers have been applied. 

Next, we construct a new equivalent quantum circuit on $n \times D$ qubits as follows. We lay out an $n \times D$ square lattice of qubits in two dimensions. Each column (of $n$ qubits) is to be initialized to $\ket{l_0}$ which, without loss of generality, is the all zeros state. The first stage in the new circuit is to obtain the state $\ket{l_1}$ in the second column. This is done by applying the gates in the first layer of the original circuit in order from top to bottom on the qubits of the first column, and following each gate with a SWAP operation that brings the qubits it acted on into the second column.

Any qubit that was not acted on in the first layer of the circuit can be thought of as acted on by an identity gate. That is, it is swapped into the second column. Next, the same thing is done to implement the second layer of the circuit and swap the qubits into the third column, except the gates are implemented in order from bottom to top. Gates are implemented on successive layers alternating between bottom-to-top order, and top-to-bottom order, until all $D$ layers are complete. After the last layer there is no need for swap operations. This procedure ensures that, at the end of the computation, the last column contains state $\ket{l_D}$, which is the output from the original circuit. (See figures \ref{fig:circ}-\ref{fig:layer3} in appendix \ref{layers}.)

The modified circuit can then be simulated using a clock Hamiltonian with bounded-range interactions in two-dimensions. Recall the clock encoding \eq{pulseclock}. Here, we use the same encoding, except the number of clock qubits will equal the total number of gates of the modified circuit including the swap gates. These clock qubits can then be ``snaked'' between the layers so that each clock bit geometrically neighbors the computational qubits acted on by the gate that the clock bit corresponds to, and the hopping terms that move the single 1 among the clock qubits are also spatially local. The construction is thus local in two spatial dimensions and involves 4-body interactions (and fewer). The wavepacket propagates down the snaking path of the clock qubits, which is of length at most $nD$, which is upper bounded by $nG$ and often much smaller. (See figure \ref{fig:layout} in appendix \ref{layers}.)

\section{Compressed Clock}
\label{sec:compressed}

In the above clock-based constructions, the total number of qubits used is on the order of $n+G$, whereas the original quantum circuit acted only on $n$ qubits. Such an increase in qubit requirement is undesirable, especially in the context of limited  qubit density and signal propagation speed, as discussed in \sect{sec:conclusion}. In this section we show how to modify the encoding of the clock so that only $G^{1/r}$ clock qubits are used, at the cost of requiring $(2r+2)$-local interactions. Achieving such clock compression together with spatial locality remains an open problem.

The most efficient encoding for the clock, in terms of qubit count, would be to store the number $x \in \{0,1,\ldots,\mathcal{G}\}$ as a binary number in $b = \lceil \log_2 (\mathcal{G}+1) \rceil$ qubits. However, a Hamiltonian based on this clock would involve $(b+2)$-qubit interactions, and is therefore, in most circumstances, not physically realistic. To achieve computational universality using a $k$-local Hamiltonian of the sort described in section \sect{sec:basic} one requires that the clock bit string encoding $x$ can be incremented to $x+1$ by flipping at most $k-2$ bits, and deciding whether a given clock bitstring encodes a given number $x \in \{0,1,\ldots,\mathcal{G}\}$ requires examining at most $k-2$ qubits. The ``pulse'' encoding \eq{pulseclock} is optimal in this respect: incrementing the clock requires flipping two qubits, and deciding whether a string encodes $x$ requires only examining the $x\th$ qubit. However, as noted above, the pulse encoding is highly suboptimal in terms of number of qubits required, namely $\mathcal{G}$.

We can interpolate between these extremes as follows. We take all $\binom{b}{r}$ strings of $b$ bits in which the number of ones (Hamming weight) is $r$. We number these from 0 to $\binom{b}{r}-1$. In such an encoding, the operators $\ket{x+1}\bra{x}$ and $\ket{x}\bra{x+1}$ appearing in the clock Hamiltonian act nontrivially on at most $2r$ bits, and therefore the clock Hamiltonian is $(2r+2)$-local. The number of bits needed for the clock register is the minimum $b$ such that $\binom{b}{r} \geq \mathcal{G}+1$, which scales as $b=O(\mathcal{G}^{1/r})$. We leave open the problem of finding the optimal tradeoff between locality and number of clock qubits.

To ensure that, given a quantum circuit, computing a description of the corresponding clock Hamiltonian is efficient, one must choose the numbering of the $\binom{b}{r}$ bit strings so that encoding of numbers into bitstrings and decoding of bitstrings into numbers are both efficient. That is, the map should admit polynomial-time classical encoding and decoding. The lexicographical numbering of the Hamming weight $r$ bitstrings admits simple methods for polynomial-time encoding and decoding for any $r$ as described for example in \cite{C71}.

\section{Conclusion and Open Problems}
\label{sec:conclusion}

The examples given above show that energy limitations alone do not impose an upper bound on computational clock speed, even if we restrict our attention to Hamiltonians that are spatially local and involve only 4-body interactions. Should one conclude that our universe admits unlimited computation speed, in principle? This seems unlikely. By introducing more detailed assumptions about physics, beyond just a limit on energy, one may recover limits on computational speed.

Two goals one could consider are achieving constant clock speed with asymptically shrinking energy (as is done in \sect{sec:dispersionless}) or achieving asymptotically growing clock speed with constant energy ($\Delta E$ and/or $\langle E \rangle$). This latter goal could be achieved by the constructions in this manuscript but with the Hamiltonian rescaled by an appropriate factor, namely a factor of order $G^2$ for $H^{\mathrm{FK}}$, and a factor of order $G$ for $H_{\mathrm{lin}}$. However, these rescaled Hamiltonians would then be sums of 4-body interactions each with norm scaling as $G$ or $G^2$. Whether a construction is possible achieving unbounded clock speed while keeping $\langle E \rangle$, $\Delta E$, \emph{and} the strength of the local interactions bounded remains an open question. In other words, limiting the interaction strength to $O(1)$ might be a sufficient additional physical assumption to recover a bound on computational speed.

If we keep the normalization used throughout this paper, where the interaction strengths are $O(1)$, we no longer obtain unbounded clock speed, but we still obtain unbounded ratio of clock speed to the energy scales $\langle E \rangle$ and $\Delta E$. Is this fully physically realistic? It seems the most fundamental aspect of this question is whether universal quantum computation by states of vanishing energy and energy uncertainty evolving according to time-indepentent Hamiltonians can be made fault tolerant against the influences of imperfect implementation and environmental noise. To our knowledge this is an open question.

One might also recover computational speed limits by bringing in spacetime considerations. In particular, assume a maximum speed $v$ for signal propagation, and a maximum density $\rho$ at which qubits can be packed. Then, for a computer of $n$ qubits in three-dimensional space, the distance between nearest-neighbor qubits is $\sim \rho^{-1/3}$ and the average distance between qubits is $~( n/\rho)^{1/3}$. Consequently, a two qubit gate must take time at least 
\begin{equation}
\label{tmin}
t_{\min} \sim \rho^{-1/3} v^{-1}
\end{equation}
to act on neighboring qubits and on the order of $\sim n^{1/3} \rho^{-1/3} v^{-1}$ to act on generic pairs of qubits.

The relevant limits to qubit density may appear at different scales depending on context. In a present-day practical context the limit on qubit density may be set by the atomic scale. In considering the computational complexity implications of relativistic quantum field theory, one can use consider a length scale $1/E$ as a cutoff\footnote{$\hbar c/E$ in units where explicit factors of $c$ and $\hbar$ are kept}, where $E$ is the available energy. More concretely, the quantum simulation algorithms of \cite{JLP12, JLP14, JLP14b} demonstrate (in simple examples of quantum field theories) that scattering processes involving particles with total energy $E$ can be faithfully simulated by discretizing space onto a lattice of spacing $O(1/E)$ and associating a register of logarithmically many qubits with each lattice site. This suggests that, effectively, the density of qubits accessible by experiments at energy scale $E$ is limited to $\widetilde{O}(E^d)$ in $d$ spatial dimensions. Perhaps the most fundamental limits to qubit density come from quantum gravity considerations such as the Bekenstein bound \cite{Bbound2, Bbound, Bousso02}. 

The practical limit on speed of information transmission often coincides, at least approximately, with the fundamental limit set by the speed of light. At the fundamental scale, it is thought that the maximum number of bits of entropy supportable within a region of spacetime faces a limit proportional to the surface area of the region, with the constant of proportionality approximately $10^{69}$ bits per square meter \cite{Bbound2, Bbound, Bousso02}. It may be tempting therefore, for a given $n$ to compute from the Bekenstein bound a corresponding $\rho$ and consequently a maximum clock speed via \eq{tmin} with the speed of light taking the place of $v$. However, it is not clear that this is a valid argument because in the regime where quantum gravity effects are significant the question of spatial locality may become subtle.\\
\\
\begin{minipage}{\textwidth}
\textbf{Acknowledgments:} I thank Scott Aaronson, Ning Bao, Michael Jarret, Manny Knill, Carl Miller, and Aaron Ostrander for useful discussions. I thank Ilya Bogdanov for suggesting the key idea behind \sect{sec:compressed} (via mathoverflow). This paper is a contribution of NIST, an agency of the US government, and is not subject to US copyright.
\end{minipage}

\clearpage
\appendix

\section{Example of Spatially Local Construction}
\label{layers}

\begin{figure}[ht]
\begin{center}
\includegraphics[width=0.3\textwidth]{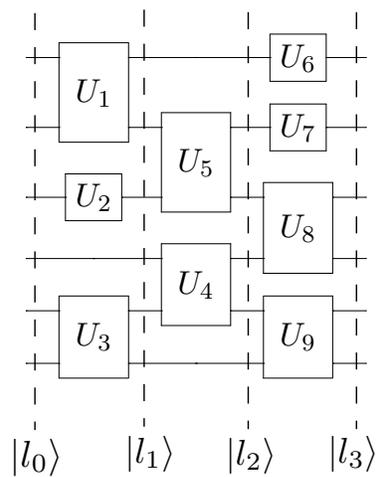}
\caption{Here is an example of a quantum circuit, referred to henceforth as ``the original circuit''. $\ket{l_j}$ labels the state after the first $j$ layers of the circuit have been applied. We have assumed that all gates in the original circuit act on nearest-neighbors in one dimension. This can always be achieved through the use of SWAP gates.\label{fig:circ}}
\end{center}
\end{figure}

\begin{figure}[ht]
\begin{center}
\includegraphics[width=0.65\textwidth]{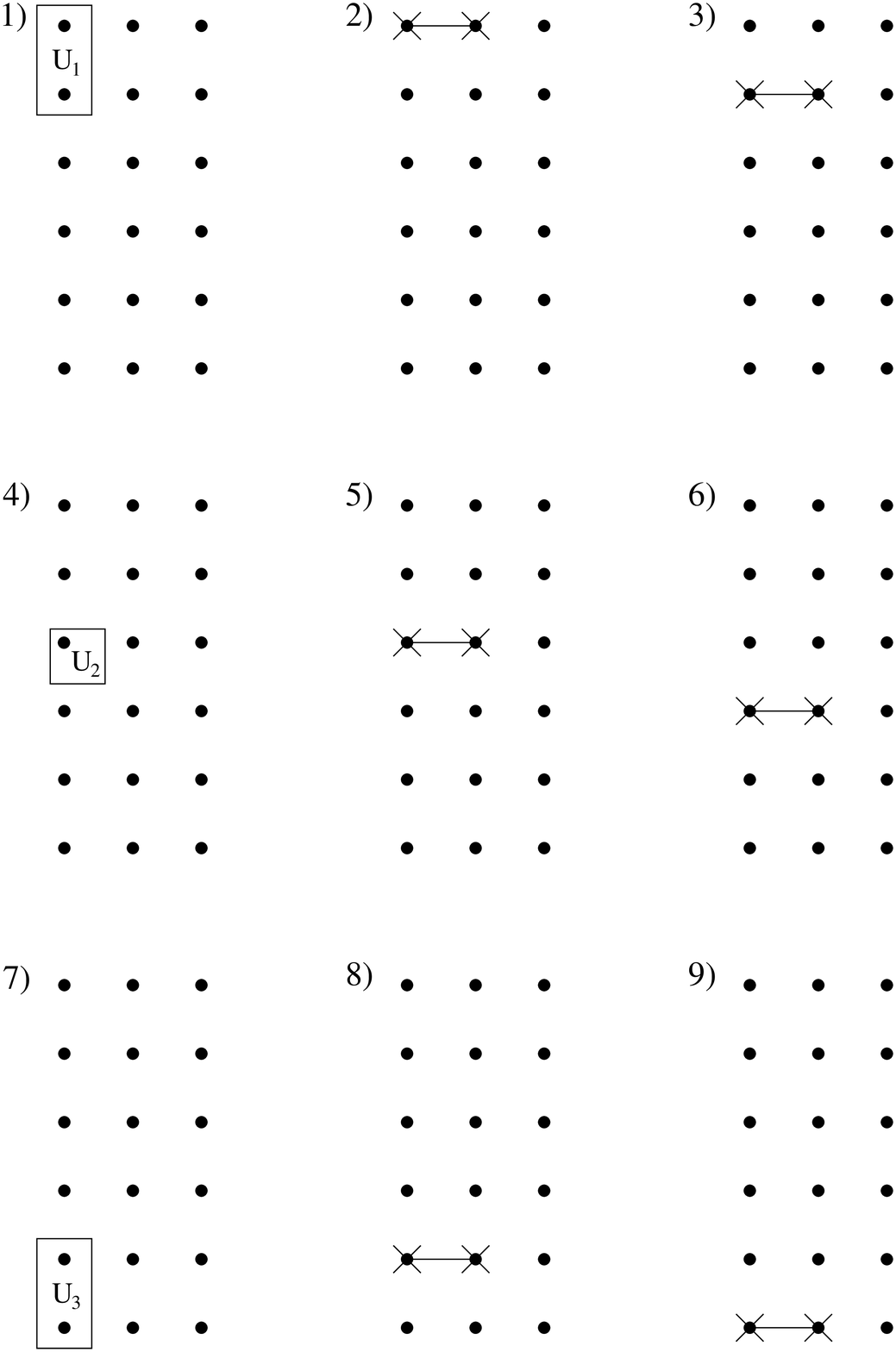}
\caption{The original circuit has depth 3 and 6 qubits. Correspondingly we have a grid of 3 columns with 6 qubits each. The above sequence of steps implements the first layer of gates. Initially each column is in the state $\ket{l_0}$, the input to the original circuit, which can be taken to be the all zeros state. After these steps are complete, the middle column contains state $\ket{l_1}$. The paired $\times$ symbols indicate swap gates. \label{fig:layer1}}
\end{center}
\end{figure}

\begin{figure}[ht]
\begin{center}
\includegraphics[width=0.65\textwidth]{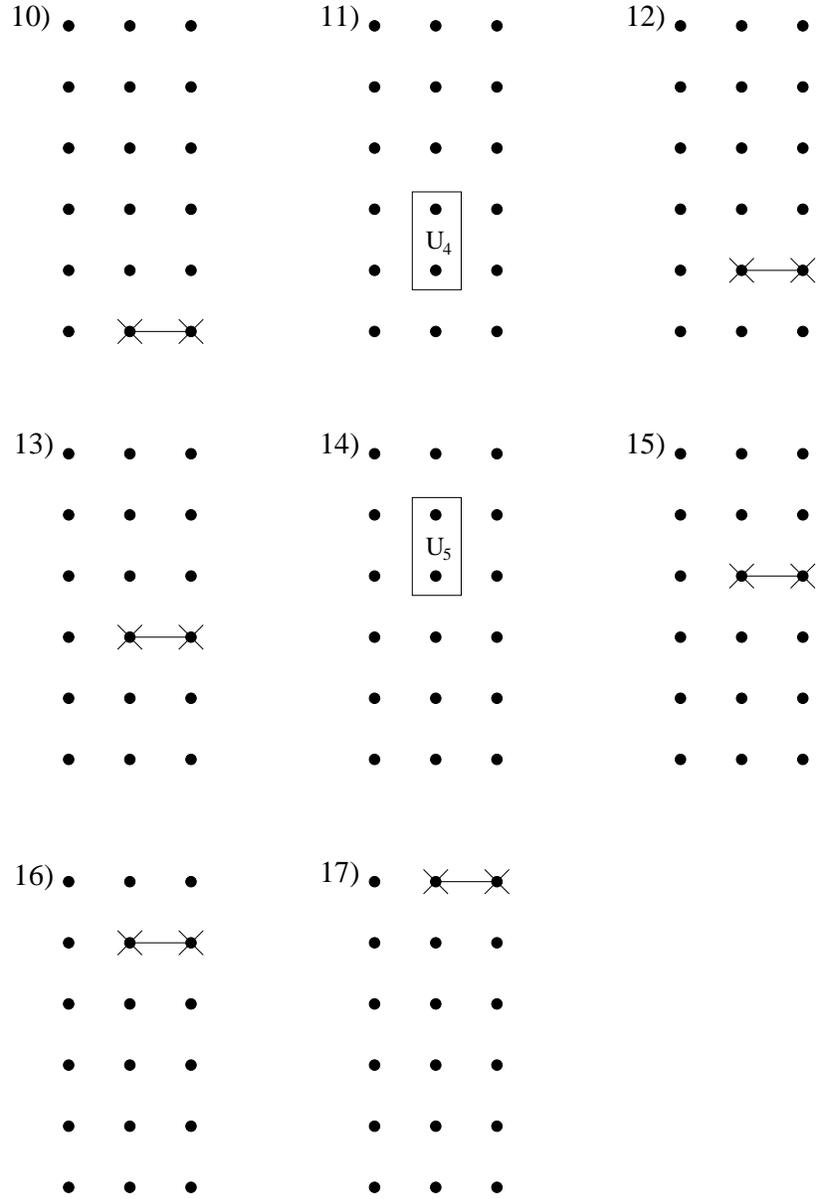}
\caption{These steps implement the second layer of gates. After these steps are complete, the right column contains state $\ket{l_2}$.
\label{fig:layer2}}
\end{center}
\end{figure}

\begin{figure}[ht]
\begin{center}
\includegraphics[width=0.65\textwidth]{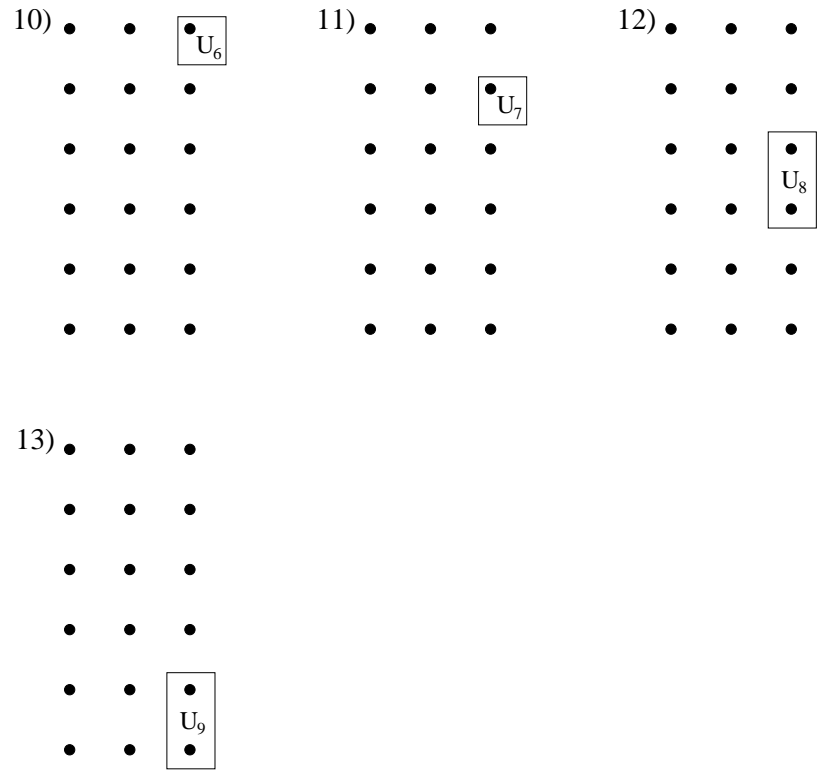}
\caption{These steps implement the third layer of gates. After these steps are complete, the right column contains state $\ket{l_3}$, which is the output of the original circuit.
\label{fig:layer3}}
\end{center}
\end{figure}

\begin{figure}[ht]
\begin{center}
\includegraphics[width=0.25\textwidth]{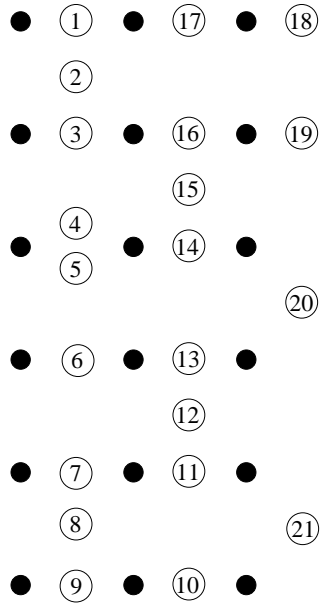}
\caption{The numbered circles represent clock qubits. The valid states of the clock are Hamming weight one strings. If the $x\th$ clock qubit is 1 then this corresponds to the $x\th$ step in figures \ref{fig:layer1}-\ref{fig:layer3}. The clock qubits are ``snaked'' among the computational qubits to ensure that the implementation of the $x\th$ step triggered by the $x\th$ clock qubit is spatially local, as is the hopping of the 1 from the $x\th$ clock qubit to the $(x+1)\th$ clock qubit.\label{fig:layout}}
\end{center}
\end{figure}

\clearpage

\bibliography{wavepacket}

\end{document}